\documentclass[preprint,preprintnumbers, prd, floatfix, superscriptaddress,nofootinbib] {revtex4-1}
\usepackage{epsfig}
\usepackage{subfigure}
\usepackage{dcolumn}% Align table columns on decimal point
\usepackage{bm}% bold math
\usepackage[usenames ,dvipsnames]{xcolor}
\usepackage{slashed}
\usepackage{graphicx,color}
\usepackage{lineno}

\begin{document}

\title{Four-body baryonic $B\to{\bf B_1\bar B'_1 B_2 \bar B'_2}$ decays}

\author{Yu-Kuo Hsiao}
\email{Email address: yukuohsiao@gmail.com}
\affiliation{School of Physics and Information Engineering,
Shanxi Normal University, Taiyuan 030031, China}

\date{\today}

\begin{abstract}
LHCb has recently reported the first observation of a four-body baryonic
$B\to{\bf B_1\bar B'_1 B_2\bar B'_2}$ decay, where
${\bf B_1\bar B'_1}$ and ${\bf B_2\bar B'_2}$ represent the two pairs of octet baryon states.
In our classification, the measured $\bar B^0\to p\bar p p\bar p$ decay is a tree dominated process 
via internal $W$-boson emission, whose branching fraction is explained as small as $2.2\times 10^{-8}$. 
We investigate for the first time the phenomenology of
other tree and penguin dominated $B\to{\bf B_1\bar B'_1 B_2\bar B'_2}$ decays, and
predict the presence of a double threshold effect, manifested as two peaks around 
$m_{\bf B_1\bar B'_1}\sim m_{\bf B_1}+m_{\bf\bar B'_1}$ and 
$m_{\bf B_2\bar B'_2}\sim m_{\bf B_2}+m_{\bf\bar B'_2}$ 
in the invariant mass spectra of ${\bf B_1\bar B'_1}$ and ${\bf B_2\bar B'_2}$, respectively.
Moreover, we predict the following branching fractions:
${\cal B}(B^-\to n\bar p p\bar p)=(1.7^{+0.4}_{-0.2}\pm 0.1^{+0.7}_{-0.4})\times 10^{-7}$,
${\cal B}(B^-\to \Lambda\bar p p\bar p)=(7.4^{+0.6}_{-0.2}\pm 0.03^{+3.6}_{-2.6})\times 10^{-7}$,
and ${\cal B}(\bar B^0_s\to \Lambda\bar \Lambda p\bar p)=(1.9^{+0.3}_{-0.1}\pm 0.01^{+1.1}_{-0.6})\times 10^{-7}$,
which are accessible to experimental facilities.
\end{abstract}
%\pacs{}

\maketitle

\section{introduction}
The baryonic $B$ decays have been extensively observed~\cite{pdg}. 
A unique phenomenon in dibaryon formation, 
known as the threshold effect, has been observed in various decays, 
including $B^-\to\Lambda\bar p\gamma$~\cite{Belle:2007lbz}, 
$B^-\to p\bar p\mu^-\bar \nu_\mu$~\cite{LHCb:2019cgl}, 
$B\to{\bf B\bar B'}M$~\cite{Belle:2007lbz,Belle:2007oni}, 
and $B\to{\bf B\bar B'}MM'$~\cite{LHCb:2017obv}. 
This effect is manifested as a peak near the threshold region of 
$m_{\bf B\bar B'}\sim m_{\bf B}+m_{\bf\bar B'}$ in the dibaryon invariant mass spectrum. 
The threshold effect indicates that $\bf B\bar B'$ tends to be produced with little extra energy. 
It can be considered as an enhancing factor~\cite{Hou:2000bz,Suzuki:2006nn} 
for ${\cal B}(B^-\to p\bar p \pi^-,p\bar p \pi^-\pi^0)$ at the level of $10^{-6}$~\cite{pdg}. 
Conversely, the formation of $p\bar p$ occurring away from the threshold region 
explains the small branching fraction ${\cal B}(\bar B^0\to p\bar p)$, 
which is $(1.27\pm 0.14)\times 10^{-8}$~\cite{LHCb:2022lff}.

It is reasonable to assume that the threshold effect also exists 
in charmless $B\to{\bf B_1\bar B'_1 B_2\bar B'_2}$ decays, 
where ${\bf B_1\bar B'_1}$ and ${\bf B_2\bar B'_2}$ represent the two pairs of octet baryons. 
One might expect their branching fractions to be as large as ${\cal B}(B\to{\bf B\bar B'}M(M'))$. 
However, experimental measurements have shown the following results~\cite{BaBar:2018gjv,LHCb:2022ntm}:
\begin{eqnarray}\label{data1}
{\cal B}(\bar B^0\to p\bar p p\bar p)
&<&2.0\times 10^{-7}\,~\text{(Babar)}\,,\nonumber\\
{\cal B}(\bar B^0\to p\bar p p\bar p)
&=&(2.2\pm 0.4\pm 0.1\pm 0.1)\times 10^{-8}\,~\text{(LHCb)}\,.
\end{eqnarray}
Clearly, ${\cal B}(\bar B^0\to p\bar p p\bar p)\sim 0.01\times {\cal B}(B^-\to p\bar p \pi^-,p\bar p \pi^-\pi^0)$ 
disagrees with a naive expectation. The enhancement of branching fractions 
near the dibaryon spectra threshold needs to be carefully examined.

As the first observed four-body fully baryonic weak decay, 
$\bar B^0\to p\bar p p\bar p$ deserves close investigation. 
It is noteworthy that this decay is a tree-dominated process 
involving internal $W$-boson emission, 
where the exchanges of the two identical particle pairs $pp$ and $\bar p\bar p$ 
can lead to indistinguishable decay configurations. 
Additionally, we should consider its counterpart with external $W$-boson emission, 
as well as penguin-dominated decay processes, 
which have not yet been mentioned or measured.

In this paper, we propose to investigate the decays $B\to{\bf B_1\bar B'_1 B_2\bar B'_2}$. 
Our analysis will focus on interpreting the branching fraction ${\cal B}(\bar B^0\to p\bar p p\bar p)$ 
to demonstrate the validity of our theoretical approach. 
Additionally, we will study $B^-\to n\bar p p\bar p$, $B^-\to\Lambda\bar p p\bar p$, 
and $\bar B^0_s\to \Lambda\bar \Lambda p\bar p$ as representative decay channels. 
We will derive the invariant mass spectra of ${\bf B_1 \bar B'_1}$ and ${\bf B_2 \bar B'_2}$. 
Our aim through this study is to contribute to the improvement of theoretical understanding 
regarding baryon-pair hadronization in weak interactions.

\section{Formalism}
%=======================
\begin{figure}[t]
\centering
\includegraphics[width=3.0in]{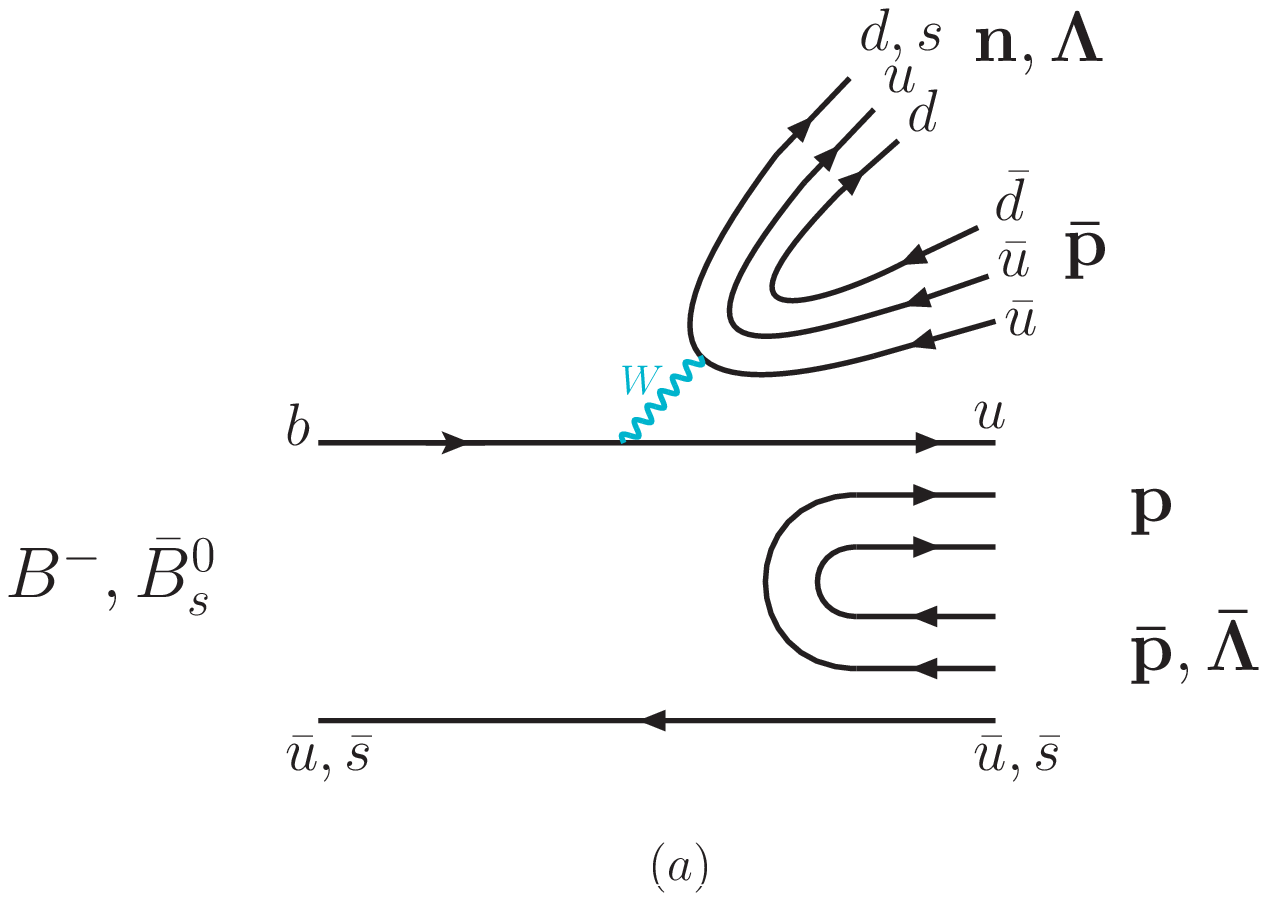}
\includegraphics[width=3.1in]{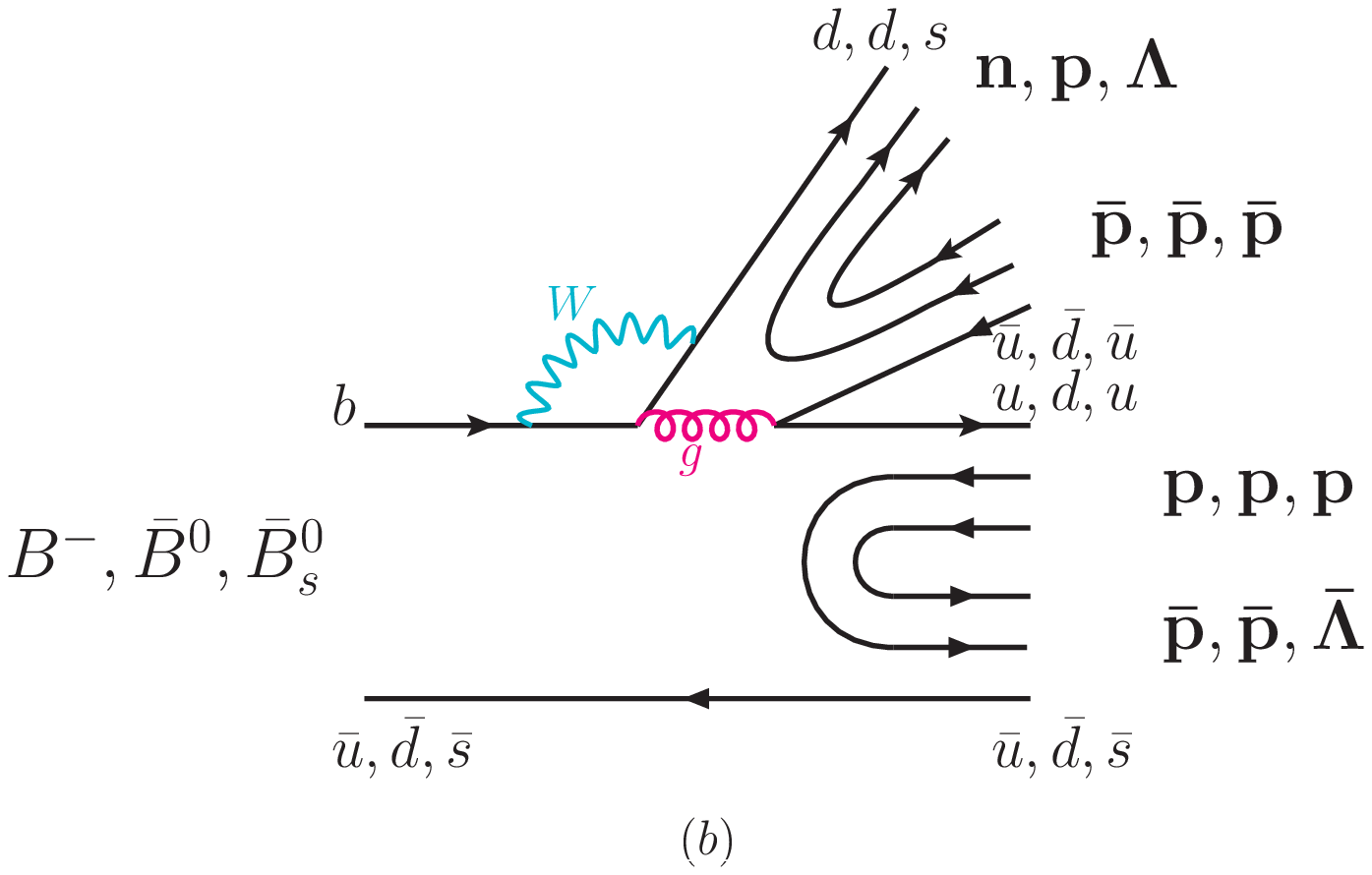}
\includegraphics[width=3.0in]{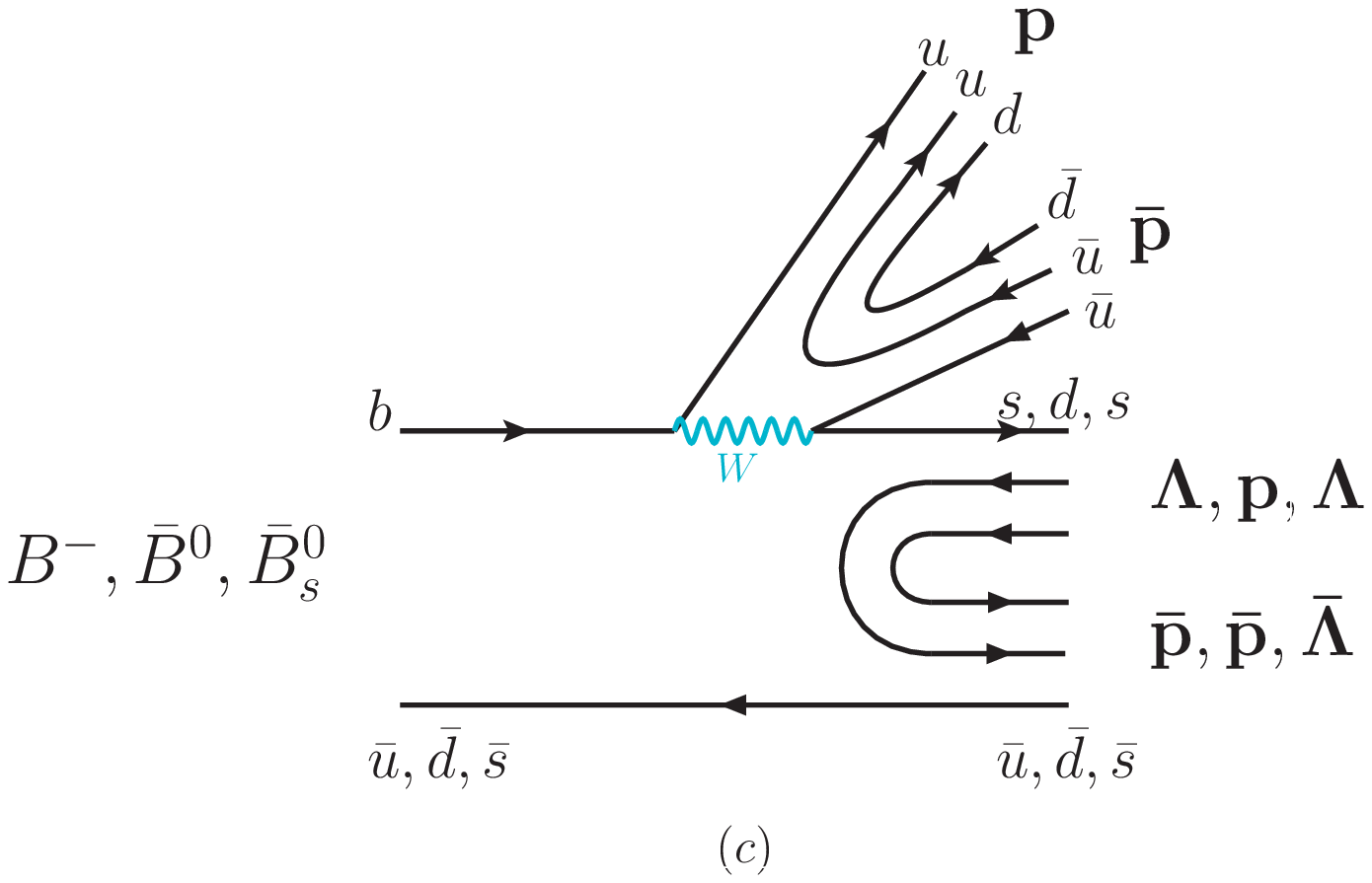}
\includegraphics[width=3.1in]{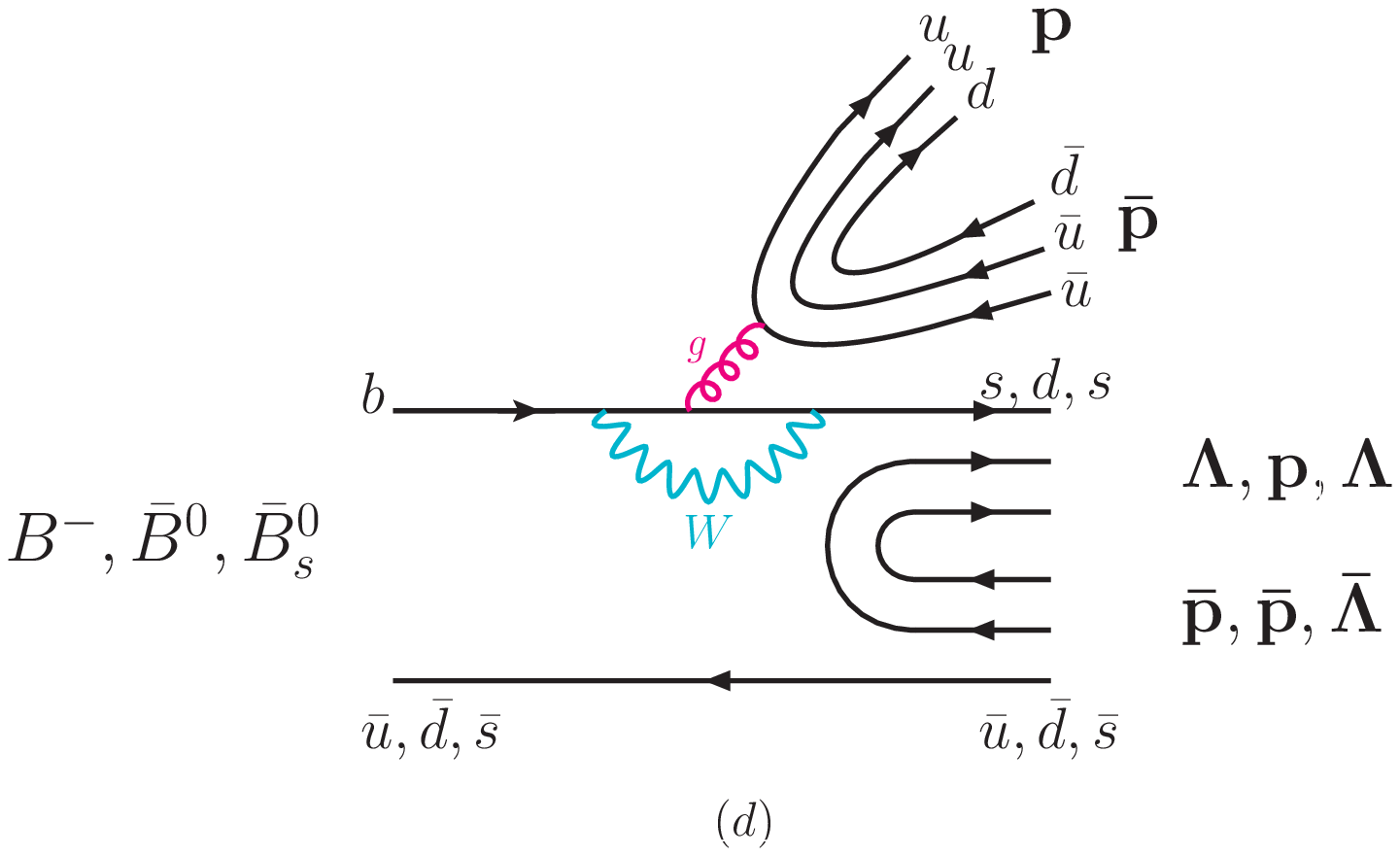}
\caption{Feynman diagrams used to illustrate 
the four-body baryonic $B\to{\bf B_1\bar B'_1\bf B_2\bar B'_2}$ decays, where:
$(a)$ depicts $(B^-,\bar B^0_s)\to (n\bar p p\bar p,\Lambda\bar p p\bar \Lambda)$, 
$(b)$ $(B^-,\bar B^0,\bar B^0_s)\to (n\bar p p\bar p,p\bar p p\bar p,\Lambda\bar p p\bar \Lambda)$, and
$(c,d)$ show $(B^-,\bar B^0,\bar B^0_s)\to p\bar p (\Lambda\bar p,p\bar p,\Lambda\bar \Lambda)$.}\label{fig1}
\end{figure}
%======================
%
Due to flavor conservation in dibaryon formation, 
we present four different configurations as shown in Fig.~\ref{fig1} 
for charmless $B\to{\bf B_1\bar B'_1 B_2\bar B'_2}$ decays. 
As an example, the decay depicted in Fig.~\ref{fig1}$(a)$ 
proceeds through the emission of an external $W$-boson, 
producing the ${\bf B_1\bar B'_1}$ pair, along with the $B$ meson transition to ${\bf B_2\bar B'_2}$. 
This configuration corresponds to the amplitude 
${\cal M}\propto \langle{\bf B_1\bar B'_1}|J^{1,\mu}|0\rangle 
\langle {\bf B_2\bar B'_2}|J^2\mu|B\rangle $ 
in the factorization approach~\cite{ali}, 
where $J^{1\,\mu}$ and $J^2_\mu$ are the currents associated with the $b\to u\bar u d$ weak decay. 
Given that the matrix elements for the vacuum $(0)\to{\bf B_1\bar B'_1}$ production 
and the $B\to{\bf B_2\bar B'_2}$ transition 
have been studied in other baryonic $B$ decays, 
the amplitude can be computed. 
Therefore, it is possible that $B\to{\bf B_1\bar B'_1 B_2\bar B'_2}$ decays 
are not so complicated as other four-body $b$-hadron processes.

\newpage
We find that $B^-\to n\bar p p\bar p$ and $\bar B^0\to p\bar p p\bar p$ 
are currently the most measurable tree-dominated processes, 
making them the typical decays to explore.
Using the effective Hamiltonian for quark-level $b$ decays~\cite{Buras:1998raa}, 
we derive the amplitudes for $B^-\to n\bar p p\bar p$ and $\bar B^0\to p\bar p p\bar p$ 
within the factorization approach~\cite{ali}. They are given by
\begin{eqnarray}\label{amp1}
{\cal M}(B^-\to n\bar p p\bar p)
&=&\frac{G_F}{\sqrt 2}\bigg\{(\alpha_1^d+\alpha_4^d)\langle n\bar p|(\bar d u)_{V-A}|0\rangle
\langle p\bar p|(\bar u b)_{V-A}|B^-\rangle\nonumber\\
&&+\alpha_6^d\langle n\bar p|(\bar d u)_{S+P}|0\rangle
\langle p\bar p|(\bar u b)_{S-P}|B^-\rangle\bigg\}\,,
\nonumber\\
{\cal M}(\bar B^0\to p\bar p p\bar p)&=&
\frac{G_F}{\sqrt 2}\bigg\{\bigg[\langle p\bar p|\alpha_+^d(\bar u u)_V-\alpha_-^d(\bar u u)_A|0\rangle+
\langle p\bar p|\beta_+^d(\bar dd)_V-\beta_-^d(\bar dd)_A|0\rangle
\nonumber\\&&
+(\alpha_4^d-\alpha_{10}^d/2)\langle p\bar p|(\bar dd)_{V-A}|0\rangle\bigg]
\langle p\bar p|(\bar d b)_{V-A}|\bar B^0\rangle
\nonumber\\&&
+\alpha_6^d\langle p\bar p|(\bar dd)_{S+P}|0\rangle
\langle p\bar p|(\bar db)_{S-P}|\bar B^0\rangle\bigg\}\,,
\end{eqnarray}
where $G_F$ is the Fermi constant, and 
we define $(\bar q_1 q_2)_{V(A)}\equiv\bar q_1 \gamma_\mu(\gamma_5) q_2$ 
and $(\bar q_1 q_2)_{S(P)}\equiv\bar q_1(\gamma_5) q_2$. 
In dealing with amplitudes involving pairs of identical particles, 
we follow the studies of $\pi^0(K_L)\to e^+ e^- e^+ e^-$~\cite{Miyazaki:1973wmu,Zhang:1997et} 
and four-body leptonic $B$ decays~\cite{Dincer:2003zq,Ivanov:2022uum}. 
For the first time, both $0\to\bf B\bar B'$ production and $B\to{\bf B\bar B'}$ transition occur in a single decay. 
It is also the first time that each of the $V^{(b)}$, $A^{(b)}$, $S^{(b)}$, 
and $P^{(b)}$ currents can form its own baryon-pair, 
which requires considering the interfering effects in the calculation.

For the penguin-dominated decays depicted in Figs.~\ref{fig1}$(b, d)$, 
the typical decays $B^-\to \Lambda\bar p p\bar p$ and 
$\bar B^0_s\to \Lambda\bar \Lambda p\bar p$ have the following amplitudes:
\begin{eqnarray}\label{amp2}
{\cal M}(B^-\to\Lambda\bar p p\bar p)&=&
\frac{G_F}{\sqrt 2}\bigg\{\bar {\cal M}_1(B^-\to\Lambda\bar p p\bar p)
+\bar {\cal M}_2(B^-\to\Lambda\bar p p\bar p)\bigg\}\,,\nonumber\\
\bar {\cal M}_1(B^-\to\Lambda\bar p p\bar p)&=&
[\langle p\bar p|\alpha_+^s(\bar u u)_V-\alpha_-^s(\bar u u)_A|0\rangle+
\langle p\bar p|\beta_+^s(\bar dd)_V-\beta_-^s(\bar dd)_A|0\rangle]\nonumber\\
&&\times\langle \Lambda\bar p|(\bar s b)_{V-A}|B^-\rangle\,,\nonumber\\
\bar {\cal M}_2(B^-\to\Lambda\bar p p\bar p)&=&
(\alpha_1^s+\alpha_4^s)\langle \Lambda\bar p|(\bar s u)_{V-A}|0\rangle
\langle p\bar p|(\bar u b)_{V-A}|B^-\rangle\nonumber\\
&&+\alpha_6^s\langle \Lambda\bar p|(\bar s u)_{S+P}|0\rangle
\langle p\bar p|(\bar u b)_{S-P}|B^-\rangle\,,\nonumber\\
{\cal M}(\bar B^0_s\to\Lambda\bar \Lambda p\bar p)&=&
\frac{G_F}{\sqrt 2}\bigg\{\bar {\cal M}_1(\bar B^0_s\to\Lambda\bar \Lambda p\bar p)
+\bar {\cal M}_2(\bar B^0_s\to\Lambda\bar \Lambda p\bar p)\bigg\}\,,\nonumber\\
\bar {\cal M}_1(\bar B^0_s\to\Lambda\bar \Lambda p\bar p)
&=&[\langle p\bar p|\alpha_+^s(\bar u u)_V-\alpha_-^s(\bar u u)_A|0\rangle+
\langle p\bar p|\beta_+^s(\bar dd)_V-\beta_-^s(\bar dd)_A|0\rangle]\nonumber\\
&&\times \langle \Lambda\bar \Lambda|(\bar s b)_{V-A}|\bar B^0_s\rangle\,,\nonumber\\
\bar {\cal M}_2(\bar B^0_s\to\Lambda\bar \Lambda p\bar p)
&=&(\alpha_1^s+\alpha_4^s)\langle \Lambda\bar p|(\bar s u)_{V-A}|0\rangle
\langle p\bar \Lambda|(\bar u b)_{V-A}|\bar B^0_s\rangle\nonumber\\
&&+\alpha_6^s\langle \Lambda\bar p|(\bar s u)_{S+P}|0\rangle
\langle p\bar\Lambda|(\bar u b)_{S-P}|\bar B^0_s\rangle\,.
\end{eqnarray}
Based on the factorization approach,
the parameters $\alpha^q_k$ and $\beta^q_l$ are given by~\cite{Hsiao:2017nga}
\begin{eqnarray}
&&
\alpha^q_\pm=\alpha_2^q+\alpha_3^q\pm\alpha_5^q+\alpha_9^q\,,\;
\nonumber\\&&
\beta^q_\pm=\alpha_3^q\pm\alpha_5^q-\alpha_9^q/2\,,\nonumber\\
&&
\alpha_{1(2)}^q=V_{ub}V_{uq}^* a_{1(2)} \,,
\alpha_j^q=-V_{tb}V_{tq}^*a_j\;(j=3,4,5,9,10)\,,\;
\nonumber\\&&
\alpha_6^q=V_{tb}V^*_{tq}2a_6\,,
\end{eqnarray}
where $q=(d,s)$, $V_{ij}$ are the Cabibbo–Kobayashi–Maskawa (CKM) matrix elements.
In the above equation, the parameters $a_{1,2}$ and $a_{3,5,9(4,6,10)}$ 
are written as~\cite{ali}
\begin{eqnarray}\label{a12}
&&a_1=c^{eff}_1+c^{eff}_2/N_c\,,\,a_2=c^{eff}_2+c^{eff}_1/N_c\,,\,\nonumber\\
&&a_{3,5,9}=c^{eff}_{3,5,9}+c^{eff}_{4,6,10}/N_c\,,\,a_{4,6,10}=c^{eff}_{4,6,10}+c^{eff}_{3,5,9}/N_c\,,
\end{eqnarray}
where $c_{1,2, ...,6}^{eff}$ and $c^{eff}_{9,10}$ are the effective Wilson coefficients, and
$N_c$ the color number.

In Eqs.~(\ref{amp1}) and (\ref{amp2}),
the matrix elements of the $0\to{\bf B_1}(p_1) {\bf\bar B'_1}(p'_1)$ productions
have been studied in the baryonic decays of
$B^-\to \Lambda\bar p$, $D_s^+\to p\bar n$, $\bar B^0\to n\bar p D^{*+}$, 
$\bar B^0\to\Lambda\bar p D^{(*)+}$, and $B\to\Lambda\bar p \pi$, 
whose parameterizations are
given by~\cite{Chua:2002wn,Chua:2002yd,Geng:2005fh,Geng:2005wt,Hsiao:2014zza,Huang:2022oli}
\begin{eqnarray}\label{FF1}
\langle {\bf B_1\bar B'_1}|V_\mu|0\rangle
&=&\bar u\bigg[F_1\gamma_\mu
+\frac{F_2}{m_1+m'_1}i\sigma_{\mu\nu}p^\nu\bigg]v\;,\nonumber\\
%&=& \bar u\bigg\{[F_1+F_2]\gamma_\mu
%+\frac{F_2}{m_{\bf B}+m_{\bf \bar B'}}(p_{\bf \bar B'}-p_{\bf B})_\mu\bigg\}v\;,\nonumber\\
\langle {\bf B_1\bar B'_1}|A_\mu|0\rangle
&=&\bar u\bigg[g_A\gamma_\mu+\frac{h_A}{m_1+m'_1}p_\mu\bigg]\gamma_5 v\,,\nonumber\\
\langle {\bf B_1\bar B'_1}|S|0\rangle &=&f_S\bar uv\;,\;
\nonumber\\
 \langle {\bf B_1\bar B'_1}|P|0\rangle&=&g_P\bar u \gamma_5 v\,,
\end{eqnarray}
where
$V_\mu(A_\mu)\equiv\bar q\gamma_\mu(\gamma_5) q'$,
$S(P)\equiv\bar q(\gamma_5) q'$,
$p_\mu=(p_1+p'_1)_\mu$,
$u(v)$ denotes the spin-1/2 spinor of the octet (anti-)baryon state, and
$F_{\bf B\bar B'}=(F_1,g_A,f_S,g_P)$ and $F'_{\bf B\bar B'}=(F_2,h_A)$
are the timelike baryonic form factors.

From the pQCD (perturbative Quantum Chromodynamics) counting rules~\cite{Brodsky:1973kr, 
Lepage:1979za,Lepage:1980fj, Brodsky:2003gs}, which provide a systematic framework 
for understanding the power-law behavior in the scattering process,
one derives that $F_{\bf B\bar B'}\propto (\alpha_s/s)^n$, where $n=2$ and $s\equiv (p_1+p'_1)^2$.
The running coupling constant in the strong interaction, denoted as $\alpha_s$, is defined as 
$\alpha_s=(4\pi/\beta_0)[\text{ln}(s/\Lambda_0^2)]^{-1}$~\cite{Lepage:1980fj}. 
Here, $\beta_0\equiv 11-2n_f/3$ is the $\beta$ function in the one-loop QCD calculation,
with the flavor number $n_f=3$ and the scale factor $\Lambda_0=0.3$~GeV. 
In particular, the power of $n=2$ reflects the fact that there are two gluon propagators involved, 
connecting the valence quarks in $\bf B_1\bar B'_1$. On the other hand, 
$F'_{\bf{B\bar B'}}=(F_2,h_A)$ requires an additional gluon to flip the chirality of the baryon pair,
indicating a plus one to $n$, such that
$F'_{\bf{B\bar B'}}$ is proportional to $1/s^{n+1}$~\cite{Chua:2002wn, Huang:2022oli}.
The momentum dependences of $F_{\bf B\bar B'}$ and $F'_{\bf B\bar B'}$ 
can hence be expressed as~\cite{Brodsky:1973kr,Brodsky:2003gs,Chua:2002wn,Geng:2006wz}
\begin{eqnarray}\label{timelikeF2}
&&
(F_1,g_A)=
\frac{(C_{F_1},C_{g_A})}{s^2}\ln\bigg(\frac{s}{\Lambda_0^2}\bigg)^{-\gamma}\,,\nonumber\\
&&
(f_S,g_P)=
\frac{(C_{f_S},C_{g_P})}{s^2}\ln\bigg(\frac{s}{\Lambda_0^2}\bigg)^{-\gamma}\,,\nonumber\\
&&
(F_2,h_A)=
\frac{(C_{F_2},C_{h_A})}{s^3}\ln\bigg(\frac{s}{\Lambda_0^2}\bigg)^{-\gamma'}\,,
\end{eqnarray}
with $\gamma^{(\prime)}=2.148(3.148)$.

%\subsection{$B\to{\bf B_2}(p_2){\bf\bar B'_2}(p'_2)$ transition form factors}
The matrix elements of the $B\to{\bf B_2}(p_2){\bf\bar B'_2}(p'_2)$ transitions 
are parameterized like those in the decays of $\bar B^0\to p\bar p D^{0(*)}$, $B\to p\bar p M$
with $M=(\pi,\rho,K^{(*)})$, $B^-\to p\bar p \ell\bar \nu_\ell$, and $B\to{\bf B\bar B'}\ell\bar \ell$,  
written as~\cite{Chua:2002wn,Geng:2006wz,Geng:2007cw,Chen:2008sw,
Hsiao:2016amt,Geng:2016fdw,Huang:2021qld,Hsiao:2018umx,Geng:2011tr,Geng:2012qn,Hsiao:2022uzx}
\begin{eqnarray}\label{FF2}
\langle {\bf B_2\bar B'_2}|V_\mu^b|B\rangle&=&
i\bar u[  g_1\gamma_{\mu}+g_2i\sigma_{\mu\nu}p^\nu +g_3 p_{\mu}
+g_4(p'_2+p_2)_\mu +g_5(p'_2-p_2)_\mu]\gamma_5v\,,\nonumber\\
\langle {\bf B_2\bar B'_2}|A_\mu^b|B\rangle&=&
i\bar u[ f_1\gamma_{\mu}+f_2i\sigma_{\mu\nu}p^\nu +f_3 p_{\mu}
+f_4(p'_2+p_2)_\mu +f_5(p'_2-p_2)_\mu]v\,,\nonumber\\
\langle {\bf B_2\bar B'_2}|S^b|B\rangle&=&
i\bar u[ \bar g_1\slashed {p}+\bar g_2(E'_2+E_2)
+\bar g_3(E'_2-E_2)]\gamma_5v\,,\nonumber\\
\langle {\bf B_2\bar B'_2}|P^b|B\rangle&=&
i\bar u[ \bar f_1\slashed {p}+\bar f_2(E'_2+E_2)
+\bar f_3(E'_2-E_2)]v\,,
\end{eqnarray}
where 
$V^b_\mu(A^b_\mu)\equiv\bar q\gamma_\mu(\gamma_5) b$,
$S^b(P^b)\equiv\bar q(\gamma_5) b$, and
$\hat F_{\bf B\bar B'}=(g_i,f_i,\bar g_j,\bar f_j)$ with $i=1,2, ...,5$ and $j=1,2,3$
are the $B\to{\bf B_2\bar B'_2}$ transition form factors.

Inspired by the pQCD counting rules,
one obtains $\hat F_{\bf B\bar B'}\propto1/t^m$
with $t\equiv (p_2+p'_2)^2$~\cite{Chua:2002wn,Geng:2006wz,
Geng:2007cw,Chen:2008sw,Hsiao:2016amt,Hsiao:2022uzx},
where $m=2+1$ is in accordance with the fact that 
there should be 2 gluons for attaching the valence quarks in $\bf B_2\bar B'_2$
and 1 for speeding up the spectator quark in $B$.
We thus present $\hat F_{\bf B\bar B'}$
as~\cite{Huang:2022oli,Hsiao:2022uzx}
\begin{eqnarray}\label{timelikeF3}
&&
(f_i,g_i)=\frac{(D_{f_i},D_{g_i})}{t^3}\,,\;
\nonumber\\
&&
(\bar f_j,\bar g_j)=\frac{(D_{\bar f_j},D_{\bar g_j})}{t^3}\,.
\end{eqnarray}
Notably,
the hard gluon picture of $1/s^n$ and $1/t^m$ has been utilized to explain or predict 
the threshold effect observed in the invariant dibaryon mass spectra of 
$B^-\to p\bar p\mu\bar \nu_\mu$~\cite{Hsiao:2022uzx,LHCb:2019cgl},
$e^+ e^-\to p\bar p$~\cite{CLEO:2005tiu},
and $B^-\to \Lambda\bar p K^+ K^-$~\cite{Hsiao:2017nga,Belle:2018ies}.

Under the $SU(2)$ helicity $[SU(2)_h]$ and $SU(3)$ flavor $[SU(3)_f]$ symmetries,
the form factors can be related to one another, and then reduced. For demonstration, 
we perform the derivation to relate $F_1$ and $g_A$.
Since the timelike form factors can be seen to behave like
the spacelike ones according to the crossing symmetry, the relation of $F_1$ and $g_A$
obtained in the spacelike region can be adopted in the timelike region.
Accordingly, we recall that~\cite{Huang:2022oli}
\begin{eqnarray}\label{Gff1}
&&
\langle {\bf B}_{R+L}\bar {\bf B}'_{R+L}|R_\mu|0\rangle\sim
\langle {\bf B}_{R+L}|R_\mu|{\bf B}'_{R+L}\rangle
%\nonumber\\&&
=\bar u\bigg[\gamma_\mu \frac{1+\gamma_5}{2}F_R+\gamma_\mu \frac{1-\gamma_5}{2}F_L\bigg]u\,,
\end{eqnarray}
where $R_\mu=(V_\mu+A_\mu)/2$ is the right-handed chiral current,
the octet baryons are decomposed as the two chiral states:
$|{\bf B}^{(\prime)}_{R+L}\rangle\equiv |{\bf B}^{(\prime)}_R\rangle+|{\bf B}^{(\prime)}_L\rangle$
with $|{\bf B}'_{R(L)}\rangle$ consisting of $q_{R} q_L q_R$ $(q_L q_R q_L)$ and their reorderings,
and $F_{R,L}$ the baryonic form factors in the chiral representation. 

When $\mu=0$ is fixed in Eq.~(\ref{Gff1}), $R_\mu$ reduces to a right-handed charge density, 
which acts on a valence quark in $|{\bf B}'_{R+L}\rangle$ and transforms ${\bf B'}$ into ${\bf B}$.
At large energy transfers, such as $s=(p_1+p'_1)^2$ around a few GeV$^2$,
the chirality $R\,(L)$ can approximately be taken as the helicity $\uparrow(\downarrow)$.
Subsequently, $q_{i,R}$ with $i=(1,2,3)$ in ${\bf B}'_{R(L)}$ can be illustrated 
to have the helicity parallel (anti-parallel) $[||(\overline{||})]$ to the helicity of ${\bf B}'$. 
Thus, the right-handed charge density that acts on $q_{i,R}$ 
can be more specifically denoted as $Q_{||(\overline{||})}(i)$.
Thus, we derive the chiral form factors $F_{R,L}$ as
$(F_R,F_L)=(e^R_{||}F_{||}+e^R_{\overline{||}}F_{\overline{||}},
e^L_{||}F_{||}+e^L_{\overline{||}}F_{\overline{||}})$
with $F_{||(\overline{||})}\equiv C_{||(\overline{||})}/s^2[\ln(s/\Lambda_0^2)]^{-\gamma}$~\cite{Lepage:1979za,Hsiao:2014zza},
where $e^{R,L}_{||(\overline{||})}$ sum over the weight factors of the baryon states,
given by
\begin{eqnarray}\label{Gff2}
e^R_{||(\overline{||})}
=\Sigma_i \langle {\bf B}_R|Q_{||(\overline{||})}(i)|{\bf B}'_R\rangle\,,\;
e^L_{||(\overline{||})}
=\Sigma_i \langle {\bf B}_L|Q_{||(\overline{||})}(i)|{\bf B}'_L\rangle\,.
\end{eqnarray}
Since $\langle {\bf B}_{R+L}|R_\mu|{\bf B}'_{R+L}\rangle$ can be decomposed into
$\langle {\bf B}_{R+L}|V_\mu|{\bf B}'_{R+L}\rangle+\langle {\bf B}_{R+L}|A_\mu|{\bf B}'_{R+L}\rangle$,
it leads to
\begin{eqnarray}
&&C_{F_1}=(e^R_{||}+e^L_{||})C_{||}+(e^R_{\overline{||}}+e^L_{\overline{||}})C_{\overline{||}}\,,\;
%\nonumber\\
%&&
C_{g_A}=(e^R_{||}-e^L_{||})C_{||}+(e^R_{\overline{||}}-e^L_{\overline{||}})C_{\overline{||}}.
\end{eqnarray}
As a result,
we relate $C_{F_1}$ and $C_{g_A}$ to $C_{||}$ and $C_{\overline{||}}$
by means of the $SU(2)_h$ and $SU(3)_f$ symmetries.
Similarly, the chiral currents $R\equiv (S+P)/2$, $R^b_\mu\equiv (V^b_\mu+A^b_\mu)/2$
and $R^b\equiv (S^b+P^b)/2$ are able to relate $(f_S,g_P)$, $(f_i,g_i)$ and $(\bar f_j,\bar g_j)$, respectively.
Here, we list the relations we need in this study, 
given by~\cite{Hsiao:2014zza,Hsiao:2017nga,Geng:2016fdw}
\begin{eqnarray}\label{timelikeF4}
&&
(C_{F_1},C_{g_A},C_{f_S},C_{g_P})=
\frac{1}{3}(4C_{||}-C_{\overline{||}},4C^\prime_{||}+C^\prime_{\overline{||}},-5\bar C_{||},-5\bar C_{||}^\prime)\,,\;\;
\nonumber\\
&&
(C_{F_1},C_{g_A})=
\frac{1}{3}(5C_{||}+C_{\overline{||}},5C^\prime_{||}-C^\prime_{\overline{||}})\,,\;\;
\nonumber\\
&&
(C_{F_1},C_{g_A},C_{f_S},C_{g_P})=
\frac{1}{3}(C_{||}+2C_{\overline{||}},C^\prime_{||}-2C^\prime_{\overline{||}},\bar C_{||},\bar C_{||}^\prime)\,,\;\;
\nonumber\\
&&
(C_{F_1},C_{g_A},C_{f_S},C_{g_P})=\sqrt\frac{3}{2}(C_{||},C_{||}^\prime,-\bar C_{||},-\bar C_{||}^\prime)\,,\;
\end{eqnarray}
for $\langle n\bar p|(\bar d u)|0\rangle$,
$\langle p\bar p|(\bar u u)|0\rangle$, $\langle p\bar p|(\bar d d)|0\rangle$, and
$\langle \Lambda\bar p|(\bar s u)|0\rangle$, and~\cite{Geng:2016fdw,Hsiao:2018umx,Hsiao:2022uzx}
\begin{eqnarray}\label{D1}
&&
(D_{g_1(f_1)},D_{g_m(f_m)},D_{\bar g_1(\bar f_1)},D_{\bar g_n,\bar f_n})
=\frac{1}{3}(5D_{||}\mp D_{\overline{||}},\pm 4D_{||}^m,5\bar D_{||}\mp \bar D_{\overline{||}},\pm 4\bar D_{||}^n)
\,,\;\nonumber\\
&&
(D_{g_1(f_1)},D_{g_m(f_m)},D_{\bar g_1(\bar f_1)},D_{\bar g_n,\bar f_n})
=\frac{1}{3}(D_{||}\mp 2 D_{\overline{||}},\mp D_{||}^m,\bar D_{||}\mp 2\bar D_{\overline{||}},\mp\bar D_{||}^n)
\,,\;\nonumber\\
&&
(D_{g_1(f_1)},D_{g_m(f_m)})
=-\sqrt\frac{3}{2}(D_{||},\pm D_{||}^m)
\,,\;\nonumber\\
&&
(D_{g_1(f_1)},D_{g_m(f_m)},D_{\bar g_1(\bar f_1)},D_{\bar g_n,\bar f_n})
=\sqrt\frac{3}{2}(D_{||},\pm D_{||}^m,\bar D_{||},\pm \bar D_{||}^n)
\,,\;\nonumber\\
&&
(D_{g_1(f_1)},D_{g_m(f_m)})
=(D_{||},\pm D_{||}^m)
\,,\;%\nonumber\\
\end{eqnarray}
for $\langle p\bar p|(\bar u b)|B^-\rangle$,
$\langle p\bar p|(\bar d b)|\bar B^0\rangle$,
$\langle \Lambda\bar p|(\bar s b)|B^-\rangle$,
$\langle p\bar \Lambda|(\bar u b)|\bar B^0_s\rangle$, and
$\langle \Lambda\bar \Lambda|(\bar s b)|\bar B^0_s\rangle$, respectively,
with $m=(2,3,4,5)$ and $n=(2,3)$.

The large angular asymmetries measured in the decays
$\bar B^0\to\Lambda\bar p\pi^+$ and $B^-\to\Lambda\bar p\pi^0$ indicate that
the $SU(2)$ chiral symmetry is broken for $F_{\bf B\bar B'}$~\cite{Belle:2007lbz,Huang:2022oli}.
Therefore, in Eq.~(\ref{timelikeF4}) we define
$C_{||(\overline{||})}^\prime\equiv C_{||(\overline{||})}+\delta C_{||(\overline{||})}$ and
$\bar C_{||}^\prime\equiv \bar C_{||}+\delta \bar C_{||}$ with
$\delta C_{||(\overline{||})}$ and $\delta \bar C_{||}$
to receive the broken effects~\cite{Huang:2022oli,Hsiao:2017nga,Geng:2016fdw}.
The chiral-flip form factors $F_2$ and $h_A$ 
do not have any relations imposed by the $SU(2)_h$ symmetry. 
In the pQCD model~\cite{Belitsky:2002kj}, $F_2$ has been calculated as
$F_2=F_1/[s\ln(s/\Lambda_0^2)]$,
which agrees with the parameterization in Eq.~(\ref{timelikeF2}). 
For $h_A$, the $SU(3)_f$ symmetry can be applied, such that 
$C_{h_A}$ in Eq.~(\ref{timelikeF2}) 
is related to the $SU(3)_f$ parameters $C_{D}$ and $C_{F}$.
Hence, we obtain~\cite{Hsiao:2014zza}
\begin{eqnarray}\label{ChA}
C_{h_A}=C_D+C_F,2C_F,-(C_D+C_F), \frac{-1}{\sqrt 6}(C_D+3C_F)\,,
\end{eqnarray}
for
$\langle n\bar p|(\bar d u)|0\rangle$,
$\langle p\bar p|(\bar u u)|0\rangle$,
$\langle p\bar p|(\bar d d)|0\rangle$, and
$\langle \Lambda\bar p|(\bar s u)|0\rangle$, respectively.
%
% =======================
\begin{figure}[t!]
\centering
\includegraphics[width=3.8in]{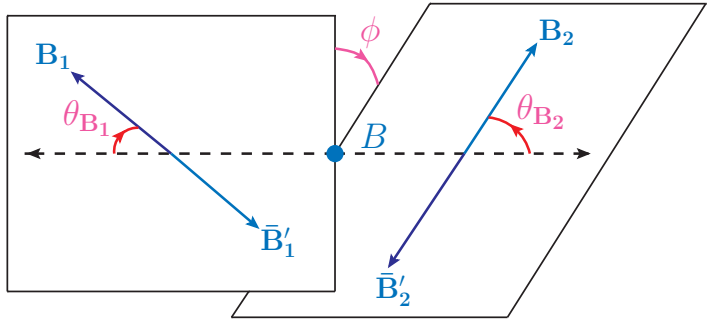}
\caption{The three angular variables $\theta_{\bf B_1,B_2}$ and $\phi$
depicted for the four-body baryonic $B\to{\bf B_1\bar B'_1\bf B_2\bar B'_2}$ decay.}\label{fig2}
\end{figure}
%========================
%
The four-body decays $B\to {\bf B_1}{\bf\bar B'_1}{\bf B_2}{\bf\bar B'_2}$
require five kinematic variables in the phase space:
$(s,t)$ and $(\theta_{\bf B_1},\theta_{\bf B_2},\phi)$~\cite{Kl4,Wise}.
The three angles $(\theta_{\bf B_1},\theta_{\bf B_2},\phi)$ are depicted in Fig.~\ref{fig2},
where $\theta_{\bf B_1(B_2)}$ is the angle between the moving directions of
${\bf B_1}$ and the ${\bf B_1\bar B'_1}$-pair
($\bf B_2$ and the ${\bf B_2\bar B'_2}$-pair)
in the $B$ meson rest frame. Moreover, $\phi$ is the angle between
the $\bf B_1\bar B'_1$ plane and the $\bf B_2\bar B'_2$ plane, which are made
by the momenta of the ${\bf B_1\bar B'_1}$ pair and ${\bf B_2\bar B'_2}$ pair,
respectively, in the $B$ meson rest frame.
The partial decay width can hence be written as~\cite{Geng:2011tr,Geng:2012qn}
\begin{eqnarray}\label{Gamma1}
d\Gamma=\frac{|{\cal M}|^2}{4(4\pi)^6 m_B^3}X
\alpha_{\bf B_1}\alpha_{\bf B_2}\,ds\,dt\,d\cos\theta_{\bf B_1}d\cos\theta_{\bf B_2}d\phi\,,
\end{eqnarray}
with $X=[(m_B^2-s-t)^2/4-st]^{1/2}$,
$\alpha_{\bf B_1}=\lambda^{1/2}(s,m_1^2,m^{\prime 2}_1)/s$, and
$\alpha_{\bf B_2}=\lambda^{1/2}(t,m_2^2,m^{\prime 2}_2)/t$,
and $\lambda(a,b,c)=a^2+b^2+c^2-2ab-2bc-2ca$,
where $|{\cal M}|^2$ is the squared amplitude with the summation over the spins.
For the integration, the allowed ranges of $(s,t)$ and $(\theta_{\bf B_1},\theta_{\bf B_2},\phi)$
are given by
\begin{eqnarray}
(m_1+m'_1)^2\leq &\,s\,&\leq (m_{B}-\sqrt{t})^2\,,\nonumber\\
(m_2+m'_2)^2\leq &\,t\,&\leq (m_{B}-m_1-m'_1)^2\,,\nonumber\\
0\leq \theta_{\bf B_1,B_2}\leq \pi&\,,&0\leq \phi\leq 2\pi\,.
\end{eqnarray}

%\newpage
\section{Numerical Analysis}
\begin{figure}[t]
\centering
\includegraphics[width=2.2in]{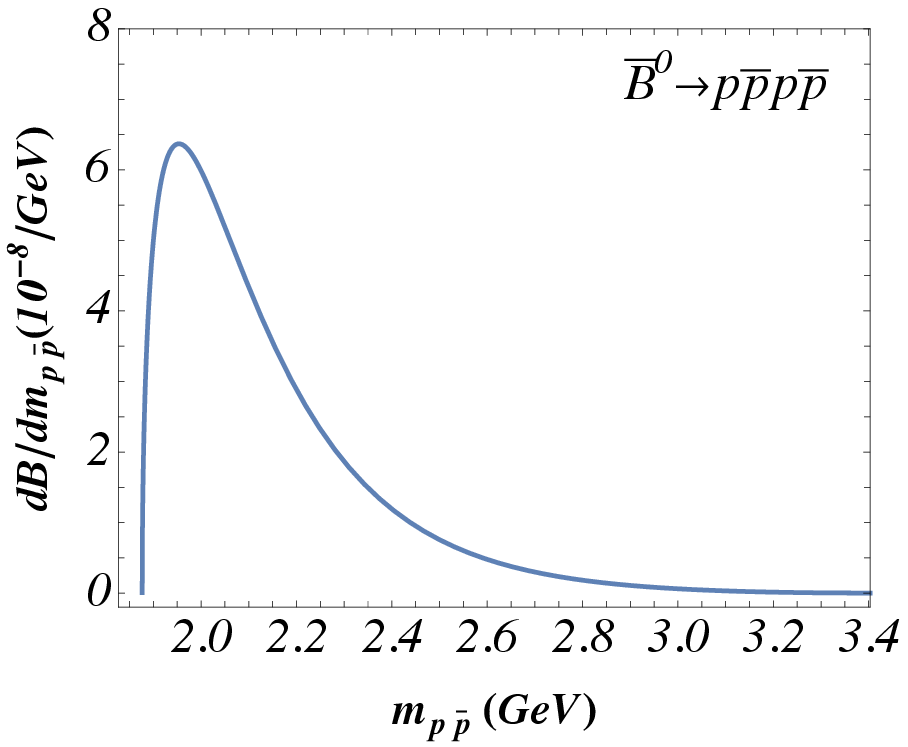}
\includegraphics[width=2.1in]{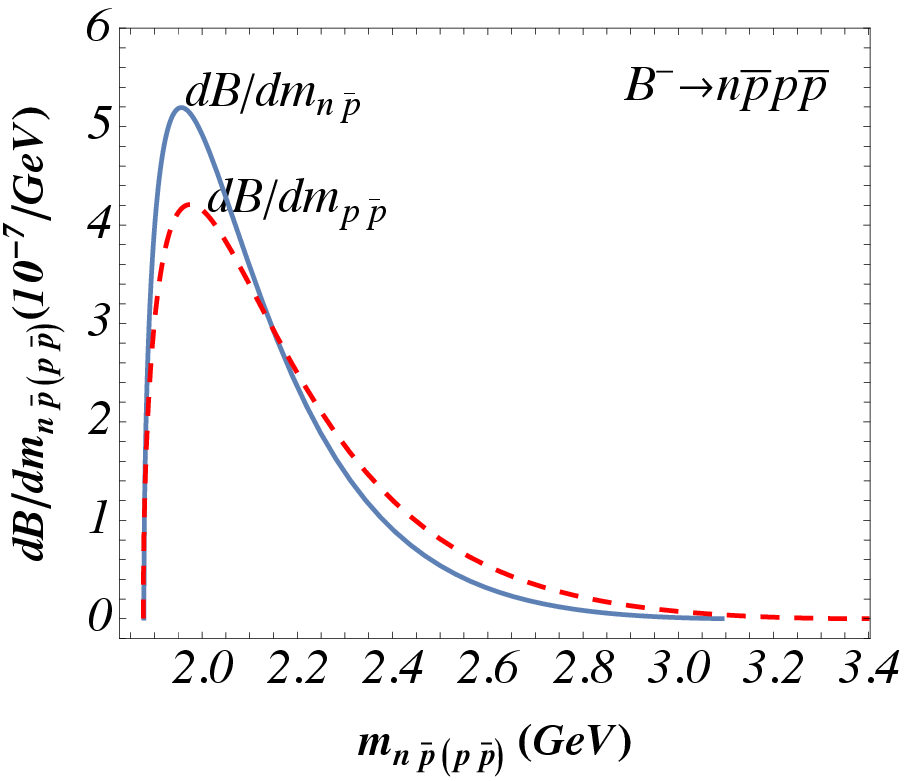}\\
\includegraphics[width=2.2in]{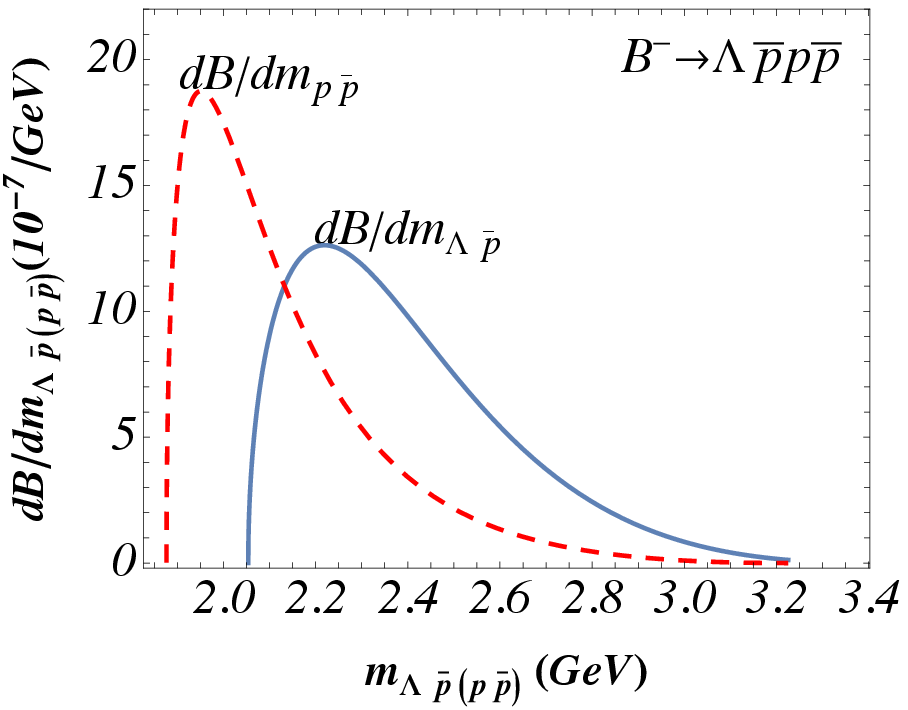}\\
\includegraphics[width=2.2in]{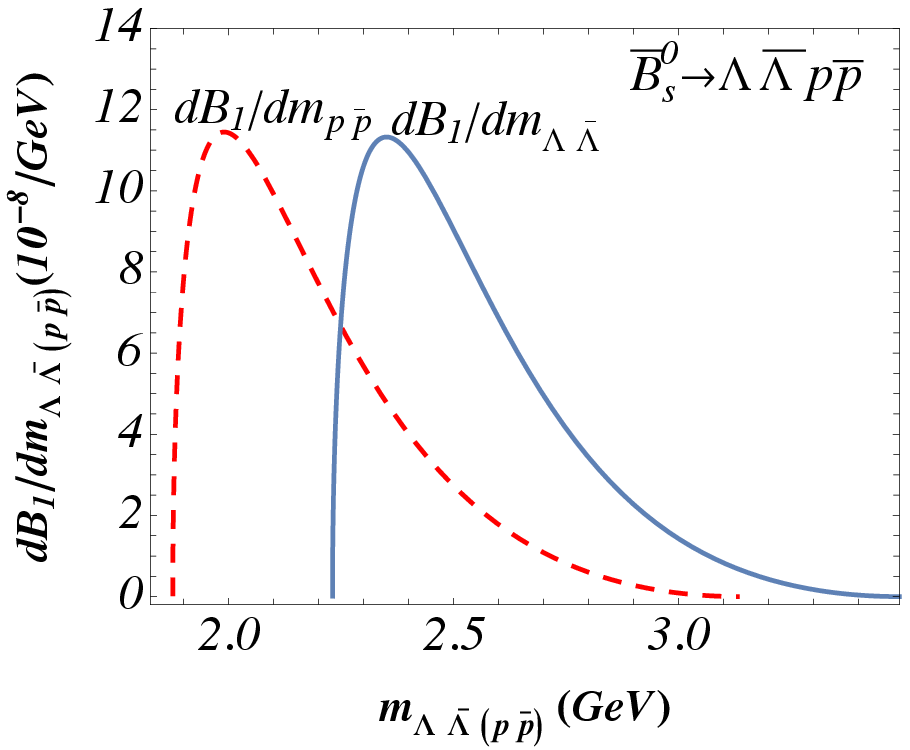}
\includegraphics[width=2.2in]{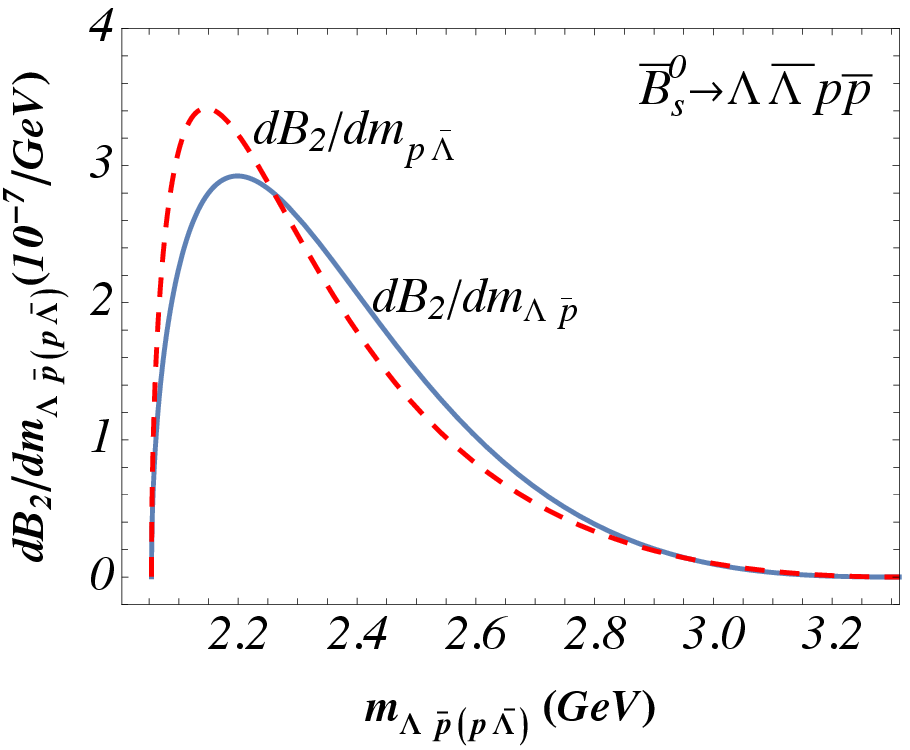}
\caption{Partial branching fractions of $B\to{\bf B_1\bar B'_1 B_2\bar B'_2}$
as a function of the relevant dibaryon invariant masses.}\label{fig4}
\end{figure}

%\newpage
%
%======================
\begin{table}[b]
\caption{The effective Wilson coefficients $c_i^{eff}$ from~\cite{ali} for the $b$ decays.}\label{effWC}
{
%\tiny
\footnotesize
\begin{tabular}{|c|cc|}
\hline
$c_i^{eff}$ &$b\to d$&$b\to s$\\
\hline\hline
$c_1^{eff}$
& $1.168$
&  $1.168$\\
$c_2^{eff}$
& $-0.365$
&  $-0.365$\\
$10^4 c_3^{eff}$
& $239.0 + 12.3i$
& $243.2 + 31.3i$\\
$10^4 c_4^{eff}$
& $-500.1 - 36.8i$
& $-512.7 - 94.0i$\\
$10^4 c_5^{eff}$
& $146.5 + 12.3i$
& $150.7 + 31.2i$\\
$10^4 c_6^{eff}$
& $-636.9 - 36.8i$
& $-649.5 - 94.0i$\\
$10^4 c_9^{eff}$
& $-112.1 -1.3i$
& $-112.3 -2.2i$\\
$10^4 c_{10}^{eff}$
& $37.5$
& $37.5$\\
\hline
\end{tabular}}
\end{table}
%======================
%
To perform the numerical analysis,
we take the four Wolfenstein parameters
$(\lambda,A,\rho,\eta)=(0.225,0.826,0.163\pm 0.010,0.357\pm 0.010)$
to parameterize the CKM matrix elements,
given by~\cite{pdg}
\begin{eqnarray}
&&
(V_{ub},V_{ud},V_{us})=(A\lambda^3(\rho-i\eta),1-\lambda^2/2,\lambda)\,,\nonumber\\
&&
(V_{tb},V_{td},V_{ts})=(1,A\lambda^3,-A\lambda^2)\,.
\end{eqnarray}
The effective Wilson coefficients $c_i^{eff}$ can be found in Table~\ref{effWC},
which take into account the quark rescattering effects in the $b$ decays~\cite{ali}.
For $F_{\bf B_1\bar B'_1}^{(\prime)}$ and $\hat F_{\bf B_2\bar B'_2}$,
the constants in Eqs.~(\ref{timelikeF4}, \ref{D1}) and (\ref{ChA}) have been extracted
in Refs.~\cite{Huang:2022oli,Hsiao:2022uzx}
to interpret the most current data in the two and three-body baryonic $B$ decays,
given by
\begin{eqnarray}\label{extraction}
&&
(C_{||},\delta C_{||},C_{\overline{||}},\delta C_{\overline{||}})=
(150.8\pm 5.7,31.9\pm 7.1,27.4\pm 27.3,-317.8\pm 169.1)\;{\rm GeV}^{4}\,,\nonumber\\
&&
(C_D,C_F)=(-761.1\pm 128.0,905.7\pm 119.8)\;{\rm GeV}^{6}\,,\nonumber\\
&&
(D_{||},D_{\overline{||}})=(11.2\pm 43.5,332.3\pm 17.2)\;{\rm GeV}^{5}\,,\nonumber\\
&&
(D_{||}^2,D_{||}^3,D_{||}^4,D_{||}^5)
=(47.7\pm 10.1,442.2\pm 103.4,-38.7\pm 9.6, 80.7\pm 27.2)\;{\rm GeV}^{4}\,,\nonumber\\
&&
(\bar D_{||},\bar D_{\overline{||}},\bar D_{||}^2,\bar D_{||}^3)
=(-59.9\pm 12.9,23.8\pm 6.8, 90.9\pm 11.1, 131.7\pm 330.7)\;{\rm GeV}^{4}\,.
\end{eqnarray}
%
% ===================
\begin{table}[b]
\caption{Our calculations for the four-body baryonic $B$ decays,
where the errors come from the non-factorizable QCD corrections, CKM matrix elements, and
the form factors, in order.}\label{results}
{
\footnotesize
%\tiny
\begin{tabular}{|l|c|c|}
\hline
$\;\;\;\;$ decay mode &our work & data\\\hline
%                    & $\ell=(e,\mu,\tau)$&\\\hline
\hline
$10^8 {\cal B}(\bar B^0\to p\bar p p\bar p)$
&$2.2\pm 0.4\pm 0.1\pm 0.4$
&$2.2\pm 0.4$~\cite{LHCb:2022ntm}\\
$10^7 {\cal B}(B^-\to n\bar p p\bar p)$
&$1.7^{+0.4}_{-0.2}\pm 0.1^{+0.7}_{-0.4}$
&-----\\
$10^7 {\cal B}(B^-\to \Lambda\bar p p\bar p)$
&$7.4^{+0.6}_{-0.2}\pm 0.03^{+3.6}_{-2.6}$
&-----\\
$10^7 {\cal B}(\bar B^0_s\to \Lambda\bar \Lambda p\bar p)$
&$1.9^{+0.3}_{-0.1}\pm 0.01^{+1.1}_{-0.6}$
&-----\\
\hline
\end{tabular}}
\end{table}
% ==============
%
Consequently, we calculate the branching fractions of
$B\to{\bf B_1\bar B'_1 B_2 \bar B'_2}$ summarized in Table~\ref{results}.
In Fig.~\ref{fig4} we draw the $m_{\bf B_1\bar B'_1}$ and $m_{\bf B_2 \bar B'_2}$ invariant mass spectra.
To get $a_i$, 
we have used the generalized edition of the factorization~\cite{ali,Hsiao:2019ann},
where $N_c^{(eff)}$, the effective color number from 2 to $\infty$,
estimates the non-factorizable QCD corrections.
According to ${\cal B}(\bar B^0\to p\bar p p\bar p)$ as measured by LHCb,
we determine $N_c^{eff}=2.50\pm 0.06$.
%
% =======================
%========================
%

%\newpage
\section{Discussion and Conclusions}
%
%=======================
%\vspace*{0.3cm}
\begin{figure}[t]
\centering
\includegraphics[width=2.4in]{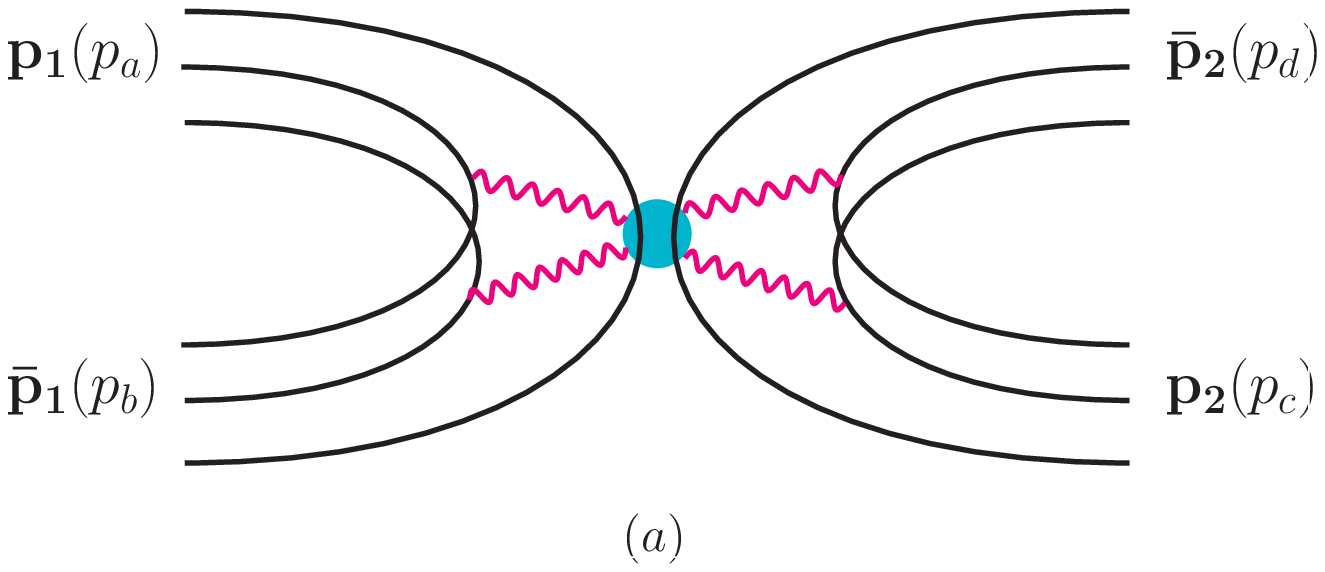}
\includegraphics[width=2.4in]{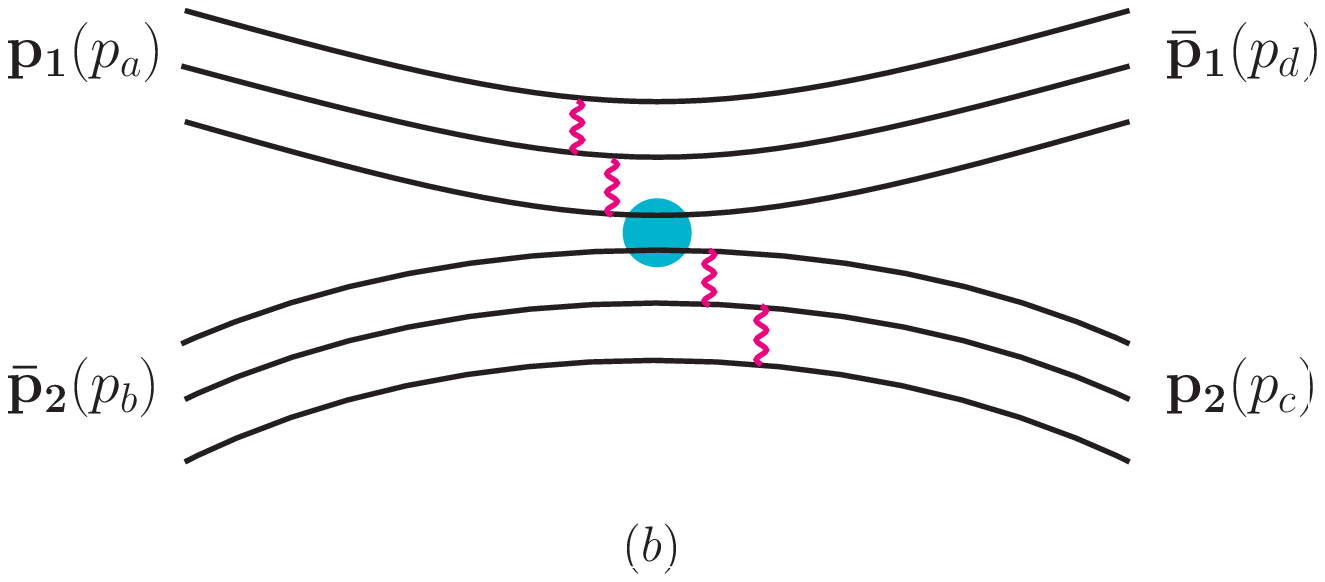}
\includegraphics[width=2.4in]{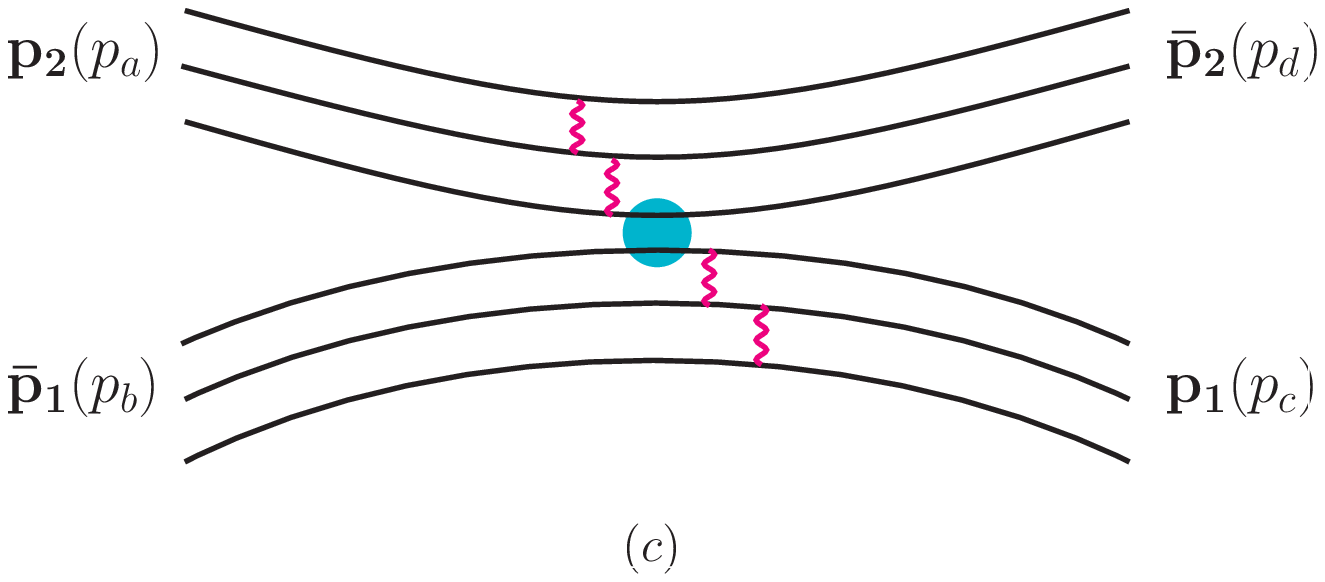}
\includegraphics[width=2.4in]{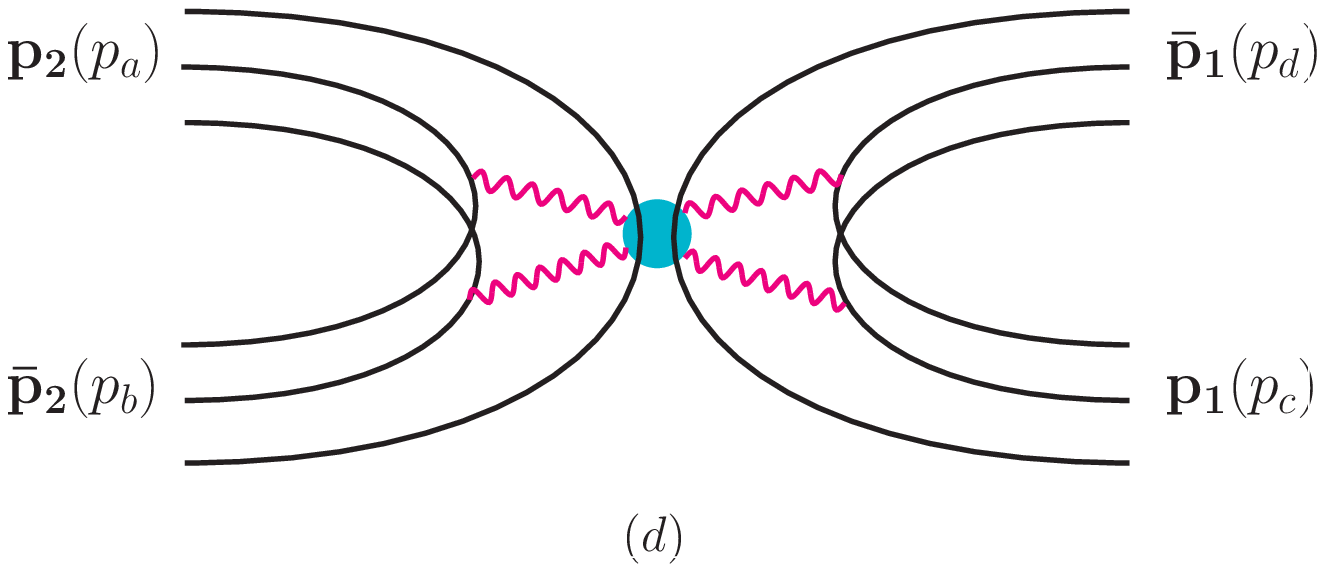}
\caption{The $\bar B^0\to p_1\bar p_1 p_2\bar p_2$ decay in the $B$ meson rest frame,
illustrating the four indistinguishable configurations 
with $m_{p_1\bar p_1}$ and $m_{p_2\bar p_2}$ around the threshold areas.
%$(a)\,{\cal M}_1$, $(b)\,{\cal M}_2$, $(c)\,{\cal M}_3$, and $(d)\,{\cal M}_4$.
}\label{fig3}
\end{figure}
%======================
%
Using the amplitude given in Eq.~(\ref{amp1}) 
and considering the baryonic form factors in Eqs.~(\ref{FF1}) and (\ref{FF2}) 
for the matrix elements of $\bf B_1\bar B'_1$ production and $B\to{\bf B_2\bar B'_2}$ transition, 
we analyze the branching fraction ${\cal B}(\bar B^0\to p\bar p p\bar p)$.
The determination of $N_c^{eff}=2.50\pm 0.06$ in the range of 2 to $\infty$ 
indicates that the generalized factorization can be successfully 
applied to our study. We also consider the indistinguishable amplitudes caused by
the identical final states in $B\to{\bf B_1\bar B'_1 B_2\bar B'_2}$.

By following the studies of $(\pi^0,K_L,\bar B^0_s)\to e^+ e^- e^+ e^-$~\cite{Miyazaki:1973wmu,
Zhang:1997et,Dincer:2003zq,Ivanov:2022uum}, the presence of identical baryon pairs $pp$ 
and $\bar p\bar p$ in $\bar B^0\to p\bar p p\bar p$ gives rise to two amplitudes. 
These are the direct amplitude ${\cal M}_{\rm dir}[\bar B^0\to p(p_1)\bar p(p'_1) p(p_2)\bar p(p'_2)]$, 
as expressed in Eq.~(\ref{amp1}), and the exchange amplitude 
${\cal M}_{\rm ex}[\bar B^0\to p(p_2)\bar p(p'_1) p(p_1)\bar p(p'2)]$, 
obtained by exchanging $p(p_1)$ and $p(p_2)$ 
in ${\cal M}_{\rm dir}[\bar B^0\to p(p_1)\bar p(p'_1) p(p_2)\bar p(p'_2)]$.
Using Eq.~(\ref{Gamma1}), the total branching fraction is divided into three parts,
${\cal B}_{\rm dir}$, ${\cal B}_{\rm ex}$, and ${\cal B}_{\rm dir\times ex}$, where
${\cal B}_{\rm dir,ex}\propto |{\cal M}_{\rm dir,ex}|^2$ and  
${\cal B}_{\rm dir\times ex}\propto {\cal M}_{\rm dir} {\cal M}_{\rm ex}^*+{\cal M}_{\rm dir}^* {\cal M}_{\rm ex}$
represents the interference term.
It is worth noting that $|{\cal M}_{\rm ex}|^2$ integrated over the total phase space is identical to
$|{\cal M}_{\rm dir}|^2$ integrated over the total phase space, 
resulting in ${\cal B}_{\rm dir}={\cal B}_{\rm ex}$.
However, calculating ${\cal B}_{\rm dir\times ex}$ can be challenging.
The difficulty arises from the momentum dependences of the form factors. 
For ${\cal M}_{\rm dir}$, we have $F_{\bf B\bar B'}\propto 1/s^2$ 
and $\hat F_{\bf B\bar B'}\propto 1/t^3$. On the other hand, 
for ${\cal M}_{\rm ex}$, we have $F_{\bf B\bar B'}^{\rm ex}\propto 1/s'^2$ 
and $\hat F_{\bf B\bar B'}^{\rm ex}\propto 1/t'^3$
with $s'\equiv (p_2+p'_1)^2$ and $t'\equiv (p_1+p'_2)$.
Fortunately, the threshold effect observed in baryonic $B$ decays 
can be utilized to estimate ${\cal B}_{\rm dir\times ex}$. 

When the threshold effect occurs, ${\cal M}_{\rm dir}$ can be associated with the configurations 
depicted in Figs.~\ref{fig3}$(a,d)$, where ${\bf B_1}{\bf\bar B'_1}$ (${\bf B_2}{\bf\bar B'_2}$) 
tends to move collinearly. In addition, 
the quark pairs can be parallel in the direction of motion~\cite{Suzuki:2006nn}, 
leading to a stronger association for their hadronization. Consequently, 
the form factors $F_{\bf B\bar B'}\propto 1/(m_1+m'_1)^4$ and $\hat F_{\bf B\bar B'}\propto 1/(m_2+m'_2)^6$ 
present in ${\cal M}_{\rm dir}$ can enhance the branching fraction~\cite{Hsiao:2018umx}.

In the case of ${\cal M}_{\rm ex}$, $\bf B_{1(2)}$ and $\bf\bar B'_{1(2)}$ 
are depicted as moving back-to-back in Figs.~\ref{fig3}$(b,c)$. 
In these configurations, each $q\bar q$ pair is anti-parallel in the direction of motion.
Since the anti-parallel configuration of the $q\bar q$ pair
requires a large energy transfer from the gluon~\cite{Hsiao:2018umx}, 
the gluon propagator is suppressed by a factor of 
$1/s'$ $(1/t')$ with $\sqrt {s'}$ $(\sqrt{t'})$ being away from the threshold. 
This mechanism has been previously employed to explain 
the suppression of ${\cal B}(\bar B^0\to p\bar p)$ to a level of $10^{-8}$~\cite{Suzuki:2006nn}, 
where the valence quark pairs also exhibit an anti-parallel configuration in the moving directions.
Specifically, we find that $F_{\bf B\bar B'}^{\rm ex}\simeq (1/4)^2 F_{\bf B\bar B'}$ and 
$\hat F_{\bf B\bar B'}^{\rm ex}\simeq (1/4)^3 \hat F_{\bf B\bar B'}$ with the occurrence of the threshold effect.
Consequently, the contribution from the interference term ${\cal B}_{\rm dir\times ex}$ 
is estimated to be $0.14\times 10^{-8}$, 
which is negligible in comparison to the experimental value in Eq.~(\ref{data1}).
Therefore, we can approximate the branching fraction  
as ${\cal B}(\bar B^0\to p\bar p p\bar p)\simeq ({\cal B}_{\rm dir}+{\cal B}_{\rm ex})/4$, 
taking into account the four indistinguishable configurations. 
This treatment can be similarly applied to other decay processes involving identical particles,
such as ${\cal B}(B^-\to n\bar p p\bar p)\simeq 
[{\cal B}_{\rm dir}(B^-\to n\bar p p\bar p)+{\cal B}_{\rm ex}(B^-\to n\bar p p\bar p)]/2$
for the two indistinguishable configurations.

The amplitude for the decay process $B\to{\bf B\bar B'}M$ can be expressed in two forms:
${\cal M}(B\to{\bf B\bar B'}M)\propto \langle {\bf B\bar B'}|(\bar q q')|0\rangle \times \langle M|(\bar q b)|B\rangle$
and ${\cal M}(B\to{\bf B\bar B'}M)\propto \langle M|(\bar q q')|0\rangle \times \langle {\bf B\bar B'}|(\bar q b)|B\rangle$.
In these expressions, the matrix elements for the production of the baryon pair ${\bf B\bar B'}$ 
are parameterized by the form factors $F_{\bf B\bar B'}$ and $\hat F_{\bf B\bar B'}$. 
These form factors are proportional to $1/s^2$ and $1/t^3$, respectively, 
where $s$ and $t$ are defined as $s,t=(p_{\bf B}+p_{\bf \bar B'})^2=m_{\bf B \bar B'}^2$. 
When $m_{\bf B \bar B'}$  is close to the threshold of $m_{\bf B \bar B'}\simeq m_{\bf B}+m_{\bf \bar B'}$,
$F_{\bf B\bar B'}$ and $\hat F_{\bf B\bar B'}$ exhibit the enhancing factors
to increase the branching fractions.

For $B\to{\bf B_1\bar B'_1 B_2\bar B'_2}$, 
the amplitude can be expressed as 
${\cal M}(B\to{\bf B_1\bar B'_1 B_2\bar B'_2})\propto 
\langle {\bf B_1\bar B'_1}|(\bar q q')|0\rangle \langle {\bf B_2\bar B'_2}|(\bar q b)|B\rangle$. 
In constrast to ${\cal M}(B\to{\bf B\bar B'}M)$,
this amplitude simultaneously involves the form factors $F_{\bf B_1\bar B'_1}\propto 1/s^2$ 
and $\hat F_{\bf B_2\bar B'_2}\propto 1/t^3$, such that
a ``double" threshold effect should arise due to
$\sqrt s=m_{\bf B_1 \bar B'_1}\simeq m_{\bf B_1}+m_{\bf \bar B'_1}$ and 
$\sqrt t =m_{\bf B_2 \bar B'_2}\simeq m_{\bf B_2}+m_{\bf \bar B'_2}$. 
This can be visualized in Fig.~\ref{fig4}.

Although the double threshold effect plays an enhancing factor
in the decay process $B\to{\bf B_1\bar B'_1 B_2\bar B'_2}$,
the fact that $\sqrt s$ and $\sqrt t$ both prefer the threshold area
imposes additional constraints on the available phase space, 
reducing the branching fraction in comparison to the case with a single threshold effect. 
Therefore, ${\cal B}(B\to{\bf B_1\bar B'_1 B_2\bar B'_2})$ 
is expected to be smaller than ${\cal B}(B\to{\bf B\bar B'}M)$.

Considering that ${\cal B}(\bar B^0\to p\bar p p\bar p)$ is found to be small, 
we search for larger branching fractions. In our classification, 
$B^-\to n\bar p p\bar p$ and $\bar B^0\to p\bar p p\bar p$ 
are tree-dominated decays with external and internal $W$ emissions, respectively. 
Additionally, the fact that $a_1>a_2$, with $a_{1,2}$ given by Eq.~(\ref{a12}), makes 
$B^-\to n\bar p p\bar p$ more favorable than $\bar B^0\to p\bar p p\bar p$.
Consequently, we obtain a larger branching fraction ${\cal B}(B^-\to n\bar p p\bar p)\simeq 1.7\times 10^{-7}$, 
which can be accessible to the Belle~II experiment. 
Notably, Belle~II has the capability to detect the elusive neutron, 
making $B^-\to n\bar p p\bar p$ a promising decay channel for experimental observation.

In the penguin-dominated decays, 
which involve both external and internal gluon emissions, 
denoted as ${\cal M}_1$ and ${\cal M}_2$ in Eq.~(\ref{amp2}), respectively, 
we calculate the branching fractions as follows: 
${\cal B}(B^-\to \Lambda\bar p p\bar p)
={\cal B}_1+{\cal B}_2$ with $({\cal B}_1,{\cal B}_2)=(1.7,5.7)\times 10^{-7}$, 
and ${\cal B}(\bar B^0_s\to \Lambda\bar \Lambda p\bar p)={\cal B}_1+{\cal B}_2$ 
with $({\cal B}_1,{\cal B}_2)=(0.6,1.3)\times 10^{-7}$. 
These results demonstrate that the two penguin configurations make compatible contributions.
In Fig.~\ref{fig4}, we illustrate the double threshold effect 
using the invariant mass spectra of $m_{\bf B_1\bar B'_1}$ and $m_{\bf B_2 \bar B'_2}$. 
Additionally, we show the four partial branching fractions as functions of 
$m_{\Lambda\bar \Lambda}$, $m_{p\bar p}$, $m_{\Lambda\bar p}$, and $m_{p\bar \Lambda}$ 
for $\bar B^0_s\to \Lambda\bar \Lambda p\bar p$. These distributions can be used to test 
whether the decay really proceeds through two distinct penguin-level configurations.

As a final remark, once our approach is validated for $B$ decays into four baryons, 
it opens up possibilities for further investigations into the direct $CP$ asymmetry and triple product asymmetry,
extensively studied in various baryonic decay processes~\cite{Hsiao:2019ann,Geng:2005wt,Geng:2006jt,Hsiao:2014mua,Sinha:2021mmx}.
Exploring these observables in the four-body fully baryonic decays 
would provide valuable insights into baryonic $CP$ violation, 
which plays a crucial role in understanding the matter-antimatter asymmetry in the universe.

In summary, we have conducted a comprehensive study of four-body fully baryonic $B$ decays, 
with a particular focus on the recently observed $\bar B^0\to p\bar p p\bar p$ decay by the LHCb collaboration. 
We have provided an explanation for its small branching fraction, where we have considered
the exchange of identical particles ($pp$ and $\bar p\bar p$) that leads to indistinguishable configurations.
Our analysis has revealed that the tree-dominated decay $B^-\to n\bar p p\bar p$ 
can be more favorable than $\bar B^0\to p\bar p p\bar p$, according to the prediction
${\cal B}(B^-\to n\bar p p\bar p)=(1.7^{+0.4}_{-0.2}\pm 0.1^{+0.7}_{-0.4})\times 10^{-7}$.
Furthermore, we have investigated penguin-dominated decay channels 
that have not been measured or studied before. Our calculation has predicted  
${\cal B}(B^-\to \Lambda\bar p p\bar p)=(7.4^{+0.6}_{-0.2}\pm 0.03^{+3.6}_{-2.6})\times 10^{-7}$ 
and ${\cal B}(\bar B^0_s\to \Lambda\bar \Lambda p\bar p)=(1.9^{+0.3}_{-0.1}\pm 0.01^{+1.1}_{-0.6})\times 10^{-7}$. 
These decay modes can be explored in future experiments such as LHCb and Belle~II.
Additionally, we have predicted the presence of a double threshold effect 
in $B\to{\bf B_1\bar B'_1 B_2\bar B'_2}$ decays.

\section*{ACKNOWLEDGMENTS}
The author would like to thank Dr.~Eduardo Rodrigues for carefully reading the manuscript
and giving valuable comments. The author would like to thank 
Profs. Jiesheng Yu, Liang Sun and Jike~Wang for useful discussions.
This work was supported by NSFC (Grants No.~11675030 and No.~12175128).

\end{document}